\documentclass[twocolumn, superscriptaddress, floatfix]{revtex4-1}
\usepackage{amsmath,graphicx}
\usepackage[pdftex,plainpages=false,colorlinks=true,linkcolor=blue, citecolor=blue, urlcolor=blue]{hyperref}
\usepackage[charter]{mathdesign}
\usepackage{bm}
\usepackage{gensymb}
\usepackage[x11names,svgnames]{xcolor}

\newcommand\rv{\mathbf{r}}

\newcommand\ev{\mathbf{e}}

\newcommand\Qv{\mathbf{Q}}

\newcommand\sv{\mathbf{s}}
\newcommand\Sv{\mathbf{S}}
\newcommand\Sigmav{\bm{\Sigma}}

\newcommand\om{\mathbf{\Omega}}
\newcommand\Gv{\mathbf{G}}
\newcommand\Tr{\mathrm{Tr}}

\newcommand\Mv{\mathbf{M}}

\begin{document}

\title{Spiral magnetism and chiral superconductivity in Kondo-Hubbard triangular lattice model}
\author{O. Ndiaye}
\affiliation{Institut de Technologie Nucléaire appliquée (ITNA), Universit\'e Cheikh Anta Diop,  5005 Dakar-Fann, Dakar, Sénégal}
\author{D. Dione}
\affiliation{Institut de Technologie Nucléaire appliquée (ITNA), Universit\'e Cheikh Anta Diop,  5005 Dakar-Fann, Dakar, Sénégal}
\author{A. Traor\'e}
\affiliation{Département de Physique, Faculté des Sciences et Techniques, Universit\'e Cheikh Anta Diop,  5005 Dakar-Fann, Dakar, Sénégal}

\author{A. S.  Ndao}
\affiliation{Institut de Technologie Nucléaire appliquée (ITNA), Universit\'e Cheikh Anta Diop,  5005 Dakar-Fann, Dakar, Sénégal}

\author{J. P. L. Faye}
\affiliation{CMC Microsystems, Sherbrooke, QC J1K 1B8, Canada} 


\date{\today}

\begin{abstract}
Building on the results of Ref. \cite{faye2018phase}, which identified an antiferromagnetic and Kondo singlet phases on the Kondo-Hubbard square lattice, we use the variational cluster approximation (VCA) to investigate the competition between these phases on a two-dimensional triangular lattice with $120^{o}$ spin orientation. In addition to the antiferromagnetic exchange interaction $J_{\perp}$ between the localized (impurity) and conduction (itinerant) electrons, our model includes the local repulsion $U$ of the conduction electrons and the Heisenberg interaction $J_H$ between the impurities. 
At half-filling, we obtain the quantum phase diagrams in both planes $(J_{\perp}, U J_{\perp})$ and $(J_{\perp}, J_{H})$. We identify a long-range, three-sublattice, spiral magnetic order which dominates the phase diagrams for small $J_{\perp}$ and moderate $U$, while a Kondo singlet phase becomes more stable at large $J_{\perp}$. The transition from the  spiral magnetic order to the Kondo singlet phase is a second-order phase transition. In the $(J_{\perp}, J_{H})$ plane, we observe that the effect of $J_H$ is to reduce the Kondo singlet phase, giving more room to the spiral  magnetic order phase. It also introduces some small magnetic oscillations of the spiral magnetic order parameter.
At finite doping and when spiral magnetism is ignored, we find superconductivity with symmetry order parameter $d+id$, which breaks time reversal symmetry. The superconducting order parameter  has a dome centered at around $5\%$ hole doping, and its amplitude decreases with increasing $J_{\perp}$. We show that spiral magnetism can coexist with $d+id$ state and that superconductivity is suppressed, indicating that these two phases are in competition. 
\end{abstract}
\maketitle
\section{Introduction}\label{Sec:Introduction}
The effects of impurities on conduction electrons is still a very active field of research. How impurities affect the motion of conduction electrons in geometrically frustrated lattices remains an interesting problem to solve. Indeed, the interaction between localized and moving electrons is crucial for many area of condensed matter physics including quantum information processing \cite{privman1998quantum, PhysRevB.78.054428, Childress281}, quantum computing \cite{steane1998quantum}, quantum materials \cite{alvarez2020doping}, and spintronics \cite{sato2002first,eschrig2015spin}. This interaction either originates $(i)$ from the hybridization of valence electrons with localized $d$ or $f$ orbitals or $(ii)$ from coupling between localized electrons and electrons spin density. In the weak hybridization regime, the first case is dominated by the Kondo exchange \cite{10.1143/PTP.32.37} interaction if the localized orbitals are slightly occupied \cite{PhysRev.149.491}. The second case corresponds to half-filled local orbitals and its low-energy physics can be understood using the so-called  Kondo lattice model. These two mechanisms describe the physics of two main families of strongly correlated heavy fermion systems which are either uranium-based or cerium-based. In uranium-based systems, the $5f$ orbitals are strongly hybridized with the $s$, $p$, or $d$ itinerant electrons, which results in strong valence charge fluctuations. Those fluctuations are frozen out in the cerium-based systems since the $4f$ level is well below the Fermi level. Just in  cerium-based systems, spin fluctuations will play a crucial role in Kondo lattice model. The latter is an effective model which is used to capture the low-energy physics of cerium-based systems  \cite{10.1143/PTP.32.37, misra2008handbook, coleman2015introduction, ortmann2015topological}. 

At half-filling, the ground state of the Kondo lattice model is an insulator. This can be explained by the formation of singlet between the conduction and impurity electrons or the presence of magnetic ordering of the impurity electrons via the Ruderman-Kittel-Kasuya-Yosida (RKKY) interaction mediated by itinerant electrons \cite{ortmann2015topological, PhysRevB.51.15630, PhysRevLett.83.796, PhysRevB.64.092406, PhysRevB.56.11820, faye2018phase}. The results obtained using mean field theory indicate an antiferromagnetic (AFM) and ferromagnetic (FM) phase order at half-filling and low doping respectively \cite{PhysRevB.20.1969}. The mean field theory shows a competition between the  magnetic order and the spin-gapped Kondo singlet phases depending on the parameters of the model. One can go from the magnetic order to the spin Kondo singlet phase by only increasing the interaction between the conduction and impurity electrons. Monte carlo simulations have been used to investigate the spin gap formation associated with hidden symmetries in a spin chain coupled with AFM Heisenberg exchange interaction \cite{PhysRevB.71.092404, PhysRevLett.100.017202, PhysRevB.82.174410}. The idea behind the structure of these spin chains, called spin-rotors, where localized spins are only coupled to conduction electrons via exchange interaction, was the study of Kondo physics \cite{PhysRevB.71.092404}.

Using the Kondo lattice model with non-interacting conduction electrons, the authors in Ref. \cite{PhysRevB.76.115108} were able to capture the qualitative physics of the heavy fermion systems on a square (bipartite) lattice. However, the Kondo lattice model fails to correctly describe the physics of those systems at lower temperature,  \cite{fulde1993model} where it is suggested that the Kondo effect must play an important role. This can be justified by the strong correlation among the charge carriers. Indeed, this strong correlation between conduction electrons  has been observed in the electron-doped cuprate $\mathrm{Nd}_{2-x}\mathrm{Ce}_{x}\mathrm{CuO}_4$ \cite{PhysRevLett.71.2481} and shown that they can enhance the Kondo temperature \cite{fulde1993model},  and therefore cannot be neglected in any Kondo lattice model at lower temperature.
On the other hand, in $\mathrm{CeAuAl_4Ge_2}$, the atomic arrangement of the cerium ions creates the conditions for geometric frustration. The essential physics of this material lies on the triangular lattice \cite{PhysRevMaterials.1.044404}. The non-interacting  Kondo lattice model on the triangular lattice has been investigated in Refs.  \cite{akagi2010spin,akagi2013ground}. One of their findings is  that a noncoplanar four-sublattice ordering emerges at and around $1/4$ filling, in addition to the $3/4$-filled case. 
To study the effect of correlations between conduction electrons at low temperature using the Kondo lattice model, a Coulomb repulsion term, with strength $U$, has been added to the Hamiltonian. This results in an Anderson-Hubbard model \cite{PhysRevLett.72.892, PhysRevB.52.R6979, PhysRevB.53.3211, PhysRevB.53.5626, PhysRevB.53.5626} which maps into an impurity model  within dynamical mean field theory \cite{PhysRevB.53.R8828, PhysRevB.54.R752, PhysRevB.56.6559}. The role of $U$ was investigated using numerical methods such as quantum Monte Carlo and bond-operator mean field theory at half-filling \cite{PhysRevB.66.045103}. The authors found that the Kondo lattice model in presence of $U$ displays a magnetic order-disorder transition and the critical Kondo interaction decreases as a function of the Hubbard repulsion.

The interplay between the quantum  magnetic order and the Kondo singlet phase on one hand, and  the competition between magnetism and superconductivity on the other hand has not been yet addressed for the Kondo-Hubbard triangular lattice model using VCA. It is well known that on a non-frustrated lattice such as the two-dimensional square (bipartite) lattice, the quantum Heisenberg model exhibits long-range N\'eel order, which is suppressed on an isotropic triangular lattice. For the latter, the classical ground state is known to have a spiral configuration in which the magnetization on each of the three sub-lattices is oriented at $120\degree$ of the other two. Results from Monte Carlo simulation of the quantum Heisenberg model on a triangular lattice show a finite sub-lattice magnetization in the ground state \cite{capriotti1999long}. However, it is well accepted that magnetism is better described by the Hubbard model. In Ref. \cite{PhysRevB.97.165110}, the superconductivity has been investigated on a triangular lattice, in both weak and strong coupling limit. Using random phase approximation, the authors identify the singlet $d+id$ pairing as the dominant state over other pairing states. Studies of the triangular-lattice $t$-$J$ model also found that $d+id$ paired state is favored among the other possibilities \cite{PhysRevLett.91.097003, PhysRevB.68.104508,PhysRevB.69.092504}. 

In this work, we investigate the interplay between quantum magnetic order and Kondo singlet phase on one hand, and the competition between magnetism and superconductivity on the other hand, within the Kondo-Hubbard triangular lattice model. In addition to the antiferromagnetic exchange interaction $J_{\perp}$ between the conduction and impurity electrons, our model includes the local repulsion $U$ of the conduction electrons, and the Heisenberg interaction $J_H$ between impurity electrons. Using the variational cluster approximation (VCA) \cite{physRevB.70.245110, faye2018phase,faye2017interplay, pavarini2014dmft, Dahnken:2004}, an approach based on the rigorous variational principal which treats short-range correlation exactly, we obtain the quantum phase diagrams in both planes $(J_{\perp}, U J_{\perp})$ and $(J_{\perp}, J_{H})$. We identify a long range, three-sublattice, spiral magnetic order which dominates the phase diagram for small $J_{\perp}$ and moderate $U$, while a Kondo singlet phase becomes more stable at large $J_{\perp}$.
At finite doping, and when spiral magnetism is ignored, we find superconductivity with symmetry order parameter $d+id$, which breaks time reversal symmetry. The superconducting order parameter  has a dome centered at around $5\%$ hole doping, and its amplitude decreases with increasing $J_{\perp}$. We show that spiral magnetism can coexist with $d+id$ state and that superconductivity is suppressed, indicating that these two phases are in competition. 

The paper is organized as follows. In Sec.~\ref{Sec:Model}, we define the model and briefly review the VCA method. We present and discuss our results in Sec.~\ref{Sec:Results}, and conclude in Sec.~\ref{conclusion}.

\section{Model and method} \label{Sec:Model}
The Kondo lattice model for heavy fermion, first introduced in Ref. \cite{PhysRevB.56.11820}, describes the motion of conduction electrons coupled to localized electrons ($f$-band). One obtains the Kondo-Hubbard model by adding the one-site Coulomb repulsion $U$.
\begin{figure}
	\centerline{\includegraphics[width=6.5cm]{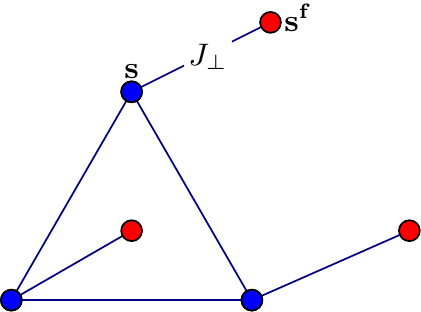}}
	\caption{The 6-site cluster used in this work.  We tile the Kondo-Hubbard triangular lattice into an infinite identical 6-site clusters. The conduction (denoted by $\Sv$) and impurity (denoted by $\Sv^f$) electrons are respectively represented by the blue and red dots. $J_\perp$ is the antiferromagnetic coupling between the conduction and impurity electrons. On each blue site, we also have the one-site Coulomb repulsion, which we do not show here.} 
	\label{fig:lattice}
\end{figure}
\subsection{Model Hamiltonian}
In this work, we consider the so-called Kondo-Hubbard triangular lattice model. The full Hamiltonian of the latter can be separated in two parts: $(i)$ the Kondo Hamiltonian part $\mathrm{H_K}$, and the Hubbard Hamiltonian part $\mathrm{H_H}$: \begin{eqnarray}\label{H} \mathrm{H} = \mathrm{H_K} + \mathrm{H_H}.\end{eqnarray} The Kondo Hamiltonian can be written as:
\begin{eqnarray}\label{kondo}
\mathrm{H_K} &=& -t\sum_{\langle i,j \rangle\sigma}c^{\dagger}_{i \sigma}c_{j \sigma}  
-\mu\sum_{i\sigma}c^{\dagger}_{i \sigma}c_{i \sigma}    
 -\mu_f\sum_{i\sigma}f^{\dagger}_{i \sigma}f_{i \sigma} \nonumber \\ &&+ J_{\perp} \sum_{i}  \sv_i   \cdot \sv^f_{i} + J_{H} \sum_{i}  \sv^f_i   \cdot \sv^f_{i}.
\end{eqnarray}
In Eq.\eqref{kondo}, $c_{i \sigma}$ $(c^{\dagger}_{i \sigma})$ and $f_{i \sigma}$ $(f^{\dagger}_{i \sigma})$ annihilate (create), respectively, a conduction and localized electron at site $i$ with spin orientation $\sigma$; likewise $\mu$ and $\mu_f$ are their chemical potentials and $t$ is the nearest-neighbor hopping amplitude. To define the model parameters in unit of the hopping amplitude, we will set $t=1$. The exchange interaction $J_\perp$, which we will assume to be an AFM coupling, couples  the  conduction spins $\sv_i = \frac{1}{2}c^{\dagger}_{i \sigma}\bm{\tau}_{\sigma, \sigma^{\prime}} c_{i \sigma^{\prime}}$ and the localized spins $\sv^f_i = \frac{1}{2}f^{\dagger}_{i \sigma}\bm{\tau}_{\sigma, \sigma^{\prime}} f_{i \sigma^{\prime}}$, where  $\bm{\tau}_{\sigma, \sigma^{\prime}}$ are the Pauli matrices. $J_H$ is a Heisenberg coupling for the localized spins which was introduced in Ref. \cite{PhysRevB.104.115103} in order to investigate the possibility of a  spin liquid phase.

The Hubbard Hamiltonian, in turn, is given by:
\begin{equation}\label{Hubbard}
    H_H = U\sum_{i}n_{i\uparrow}n_{i \downarrow} + U_f\sum_{i}n^f_{i\uparrow}n^f_{i \downarrow},
\end{equation}
 where $U$ and $U_f$ are the one-sites Coulomb repulsion for the conduction and localized electrons respectively, $n_{i\sigma} = c^{\dagger}_{i \sigma}c_{i \sigma}$ and  $n^f_{i\sigma} = f^{\dagger}_{i \sigma}f_{i \sigma}$ with $\sigma = \uparrow, \downarrow$. In our Kondo-Hubbard lattice model, the $f$-electron are truly locals. This can justified by the absence of hybridization between the conduction and localized electrons, and the fact that there is non hopping between sites occupied by the impurity electrons. In our model, to prevent hopping between these sites, we set the local Coulomb repulsion $U_f = 100t$, which will be larger than any parameter of the model. To make sure that we have exactly one $f$-electron per site, we take $\mu_f = U_f/2$.
 
In the limit of large $J_\perp \gg t$, it can be shown that the conduction and localized spins form a singlet phase, known as Kondo singlet. At half-filling, decreasing the strength of the exchange interaction $J_\perp$ leads to a spiral magnetic order phase. At very low temperature, going away from half-filling by varying the chemical potential $\mu$ of the conduction electrons, the system can transit from the Kondo singlet phase to a superconducting phase. In this paper, we will investigate these two cases.

\subsection{Variational Cluster Approximation method}\label{Sec:VCA}
The variational cluster approximation (VCA) is a quantum cluster approach which uses exact diagonalization  as a solver at zero temperature \cite{Dahnken:2004}. VCA can be viewed as an extension of cluster perturbation theory \cite{PhysRevB.66.075129} which is based on Potthoff’s self-energy functional approach \cite{potthoff2003variational, potthoff2003self}. The VCA method has been used to study the competition between magnetism and superconductivity on the two-dimensional Hubbard model of strongly correlated systems such as high-$T_c$ cuprates \cite{PhysRevLett.94.156404, PhysRevB.74.235117}. A more exposed review can be found in Ref. \cite{Potthoff:2012}. VCA starts by a tiling of the lattice into an infinite identical small clusters. In VCA, the size of the cluster must be small enough for the electron Green function to be computed numerically using exact diagonalization method. In this paper, we use a 6-site cluster composed of 3-site for the conduction band and 3-site for the localized band, as described in Fig. \ref{fig:lattice}. VCA works by distinguishing the original system defined on an infinite lattice and described by $H$ in \eqref{H}, to a reference system (cluster) described by a Hamiltonian $H^{\prime}$. To obtain $H^{\prime}$, one just removes the inter-cluster hopping in $H$ to end up with a set of small systems each governed by  $H^{\prime}$. Broken symmetries will be proved by adding their corresponding Weiss fields to $H^{\prime}$, and more generally, any one-body term can be added. Using a variational principal, we find the optimal one-body part of $H^{\prime}$. The electron self-energy ${\bm \Sigma}$, associated with $H^{\prime}$, is used as a variational self-energy in order to construct the Potthoff self-energy functional \cite{potthoff2003self}:
\begin{eqnarray}\label{functional_potthoff}
\om[\Sigmav (\xi)] =\om^{\prime}[\Sigmav(\xi)] + \Tr \ln [\frac{(\Gv^{-1}_0 -\Sigmav(\xi))^{-1}}{\Gv^{\prime}(\xi)}].
\end{eqnarray}
In Eq. \eqref{functional_potthoff}, the parameters $\Gv^{\prime}$ and $\Gv_{0}$ are the Green functions of the cluster and the non-interacting lattice respectively. The parameters that define the one-body part of $H^{\prime}$ are denoted by $\xi$, $\Tr$ is a functional trace, i.e., a sum over frequencies, momenta and bands, and $\om^{\prime}$ is the grand potential of the cluster, i.e., its ground state energy noting that the chemical potential $\mu$ is included in the Hamiltonian. $\Gv^{\prime}$ and $\om^{\prime}$ are computed numerically via the Lanczos method at zero temperature.

The Potthoff functional $\om[\Sigmav (\xi)]$ in Eq.\eqref{functional_potthoff} is computed exactly, but on a restricted space of the self-energies $\Sigmav (\xi)$ that are the physical self-energies of the reference Hamiltonian $H^{\prime}$. We use a standard optimization method (e.g. Newton-Raphson) in the space of parameters $\xi$ to find the stationary value of $\om(\xi)$:
\begin{eqnarray}\label{om}
\frac{\partial \om(\xi)}{\partial \xi} =0.
\end{eqnarray}
This represents the best possible value of the self-energy
$\Sigmav (\xi)$, which is used, together with the non-interacting Green function $\Gv_0$, to construct an approximate Green function
G for the original lattice Hamiltonian $H$. From that Green
function one can compute the average of any one-body operator, in particular the order parameters associated with
magnetism and superconductivity. The actual value of $\om(\xi)$ at the stationary point is a good approximation to the physical grand potential of the lattice Hamiltonian $H$.

There may be more than one stationary solutions to
Eq.\eqref{om}. For instance, a normal state solution in which all Weiss fields used to describe broken symmetries are zero,
and another solution, with a non-zero Weiss field, describing a broken symmetry state. As an additional principle, we
assert that the solution with the lowest value of the functional in Eq. \eqref{functional_potthoff} is the physical solution \cite{potthoff2006systematics}. Thus competing phases may be compared via their value of the grand potential $\om$, obtained by introducing different Weiss fields. 
\section{Results and discussion}\label{Sec:Results}
VCA is a method that does not require the factorization of the interaction, and more importantly, it takes into account exactly short-range correlation within the cluster. It is superior to static mean field approaches, and is more suitable for investigating broken symmetry phases such as the magnetic and superconducting order. As discussed in Sec. \ref{Sec:VCA}, the approximation originates from the limited space of the self-energies on which the variational principal is applied. However, $\Gv$ is still defined on the infinite lattice.
In Sec. \ref{hf}, we define the Weiss field for spiral magnetic order  and present the phase diagrams in the planes $(J_{\perp}, U J_{\perp})$ and $(J_{\perp}, J_{H})$ at half-filling. 
\subsection{Phase diagrams at half-filling}\label{hf}
The Weiss field for spiral magnetic order, that we will consider in this work, can be expressed as $H_h = h\hat{\Mv}$ with:
\begin{equation}\label{champ de weiss}
\hat{\Mv} =  \sum_{i \in A}\ev_A\cdot \bm \Lambda_i + \sum_{i \in B }\ev_B \cdot \bm \Lambda_i +\sum_{i \in C} \ev_C\cdot \bm \Lambda_i,
\end{equation}
where $A$, $B$ and $C$ represent the sub-lattices of the triangular lattice, the unit vectors $\ev_{A,B,C}$ are oriented at $120^{\circ}$ of each other and, $\bm \Lambda_i = \Sv_i$ and $\bm \Lambda_i = \Sv^f_i$ respectively for the conduction and localized spins. We assume that each conduction site is coupled with an impurity site.  At half-filling, we set $\mu = \frac{U}{2}$ and $\mu_f = \frac{U_f}{2}$, respectively for the conduction, and localized electrons. The spiral magnetic order parameter (SMOP) is the expectation value of the operator $\hat{\Mv}$ divided by the number of lattice sites. It can  be obtained from the lattice Green function \cite{sahebsara2008hubbard} as:
\begin{equation}\label{op}
\mathrm{SMOP} = 2i\int\frac{d^2 K}{(2\pi)^2}\int\frac{d\omega}{2\pi}\Gv_{a,b} \Mv_{a, b},
\end{equation}
where the indices $a, b$ are composite in the sense that they include both the cluster site and spin, i.e. $a = (i, \sigma)$. The frequency integral in Eq. \eqref{op} is taken along the positive imaginary axis, and the matrix $\Mv_{a, b}$ expresses $\hat{\Mv}$ as one-body term according to $\Mv_{a, b}c^{\dagger}_{a}c_{b}$. 

In the limit $J_\perp \gg t$, the ground state of in Eq. \eqref{H} can be shown to be a product of singlets, formed locally between the conduction and localized electrons. At small and moderate value of $J_\perp$, other broken symmetries can exist, like spiral magnetic ordering.
\begin{figure}
	\includegraphics[width=9.0cm]{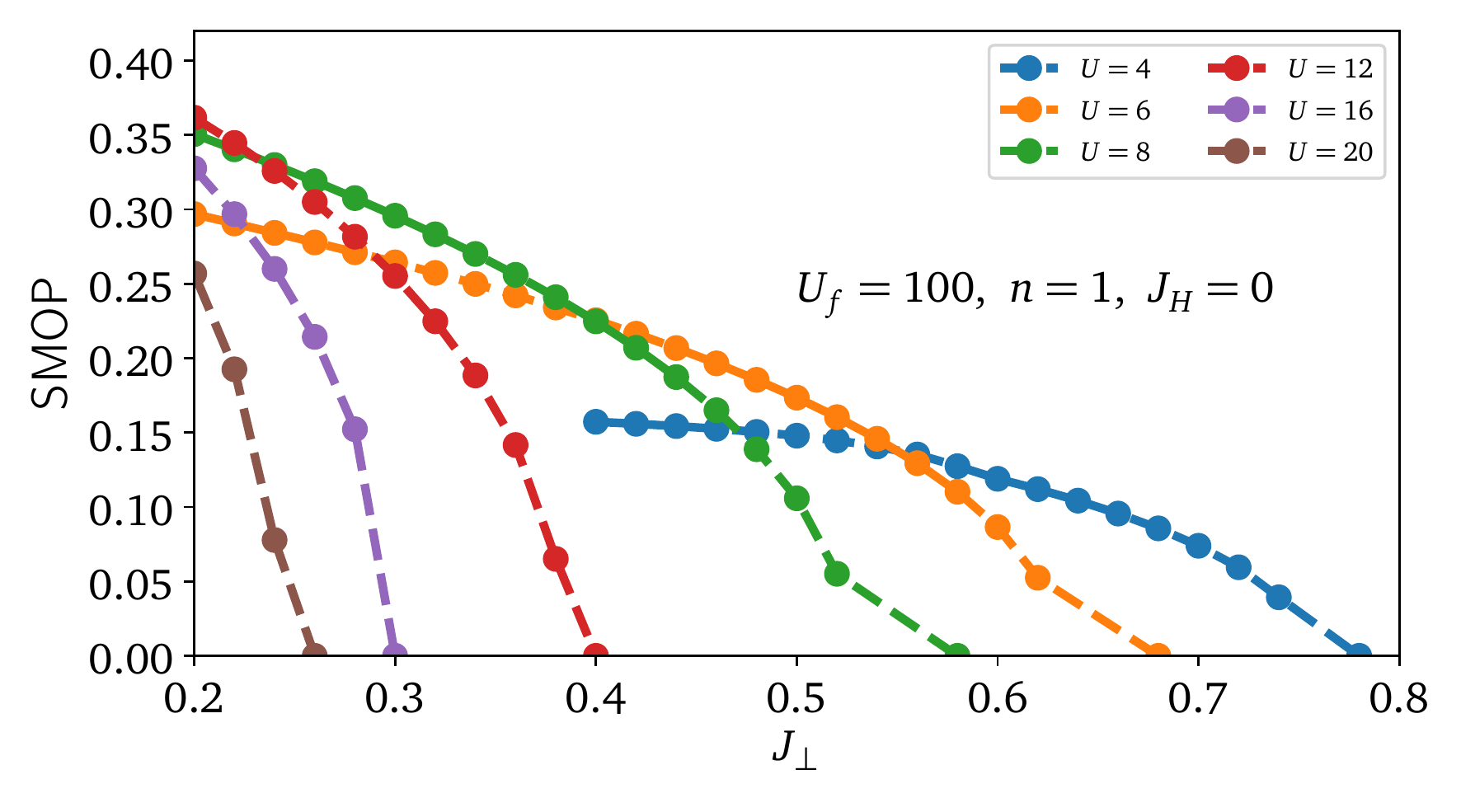}
	\centerline{\includegraphics[width=9.0cm]{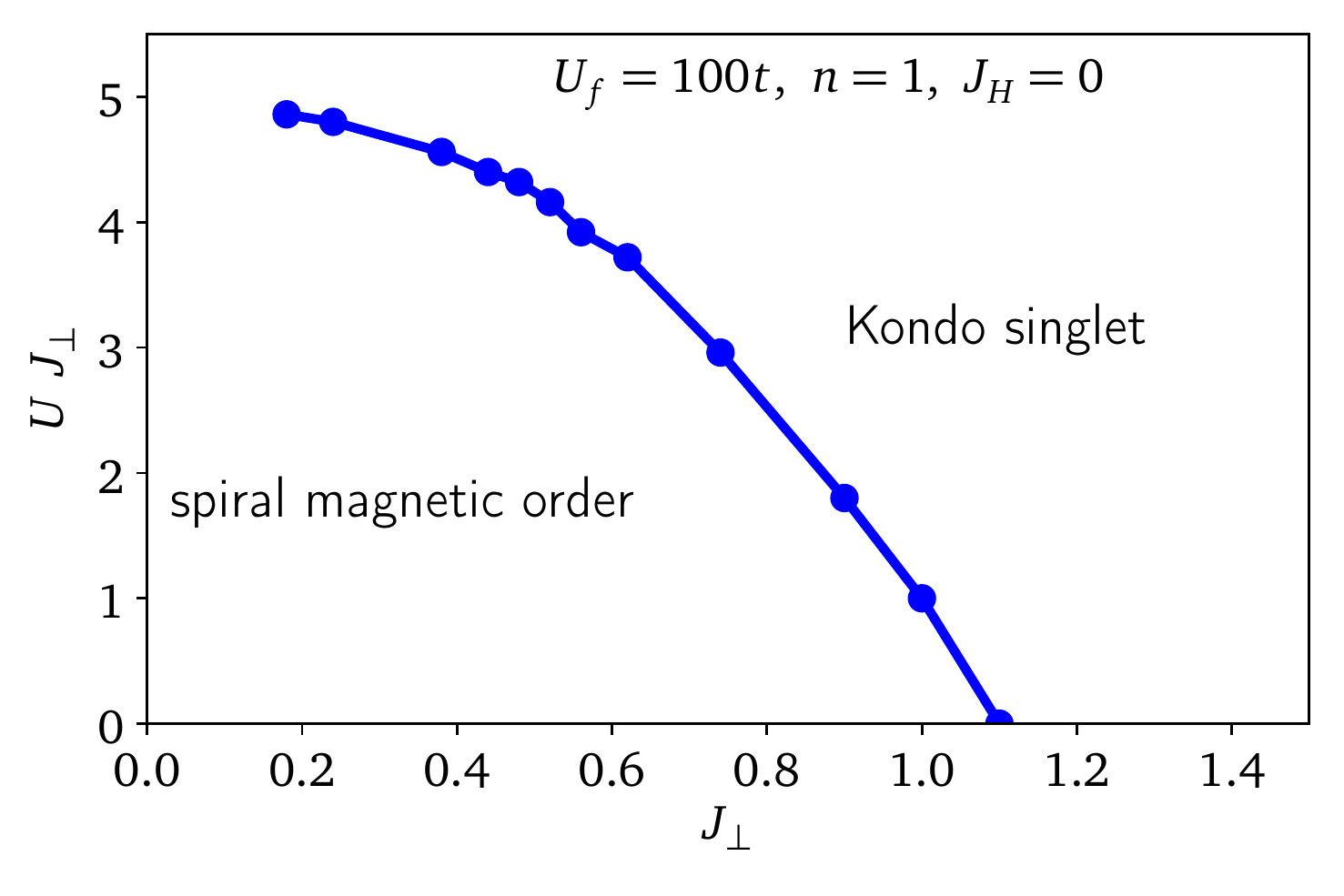}}
	\caption{Top panel: Spiral magnetic order parameter SMOP, at half-filling, in function of $J_{\perp}$ for different values of Coulomb interaction $U$ at $J_H=0$. We show the SMOP for some values of $U$. Low panel: Phase diagram at half-filling in $(J_{\perp}, U)$ plane. The critical line (in blue) is obtained by collecting the critical $J_{\perp}$ for fixed Coulomb interaction $U$. We also set $U_{f} = 100t$.} 
	\label{fig:SMOP-PD}
\end{figure}
In Fig \ref{fig:SMOP-PD} (top panel), we show the spiral magnetic order parameter SMOP in function of the coupling $J_{\perp}$ for different values of $U$. We set $U_f = 100 t$ as discussed in Sec.\ref{Sec:Model}, and $J_H = 0$. At half-filling, the number of electron by site, i.e. the electronic density is set to $n = 1$, for both the conduction and localized orbitals. We observe that the critical couplings  $J_\perp$, points where the SMOP disappears, decreases with increasing Coulomb repulsion $U$, leaving room for Kondo singlet. Thus, the Hubbard interaction $U$ reduces the spiral magnetic order phase. However, even for large $U$ ($U = 30t$ for example), the spiral magnetic order parameter still remains finite.
From Fig.~\ref{fig:SMOP-PD} (upper panel), we remark that the spiral magnetic order parameter goes to zero smoothly at critical coupling $J_\perp$. This indicates that the transition from  spiral magnetic order phase to Kondo singlet phase is a continuous (second-order) phase transition.

The phase diagram at half-filling for $J_H = 0$ is shown in Fig.~\ref{fig:SMOP-PD} (low panel) in the $(J_\perp, ~J_\perp U)$ plane.  
We obtain this quantum phase diagram  by collecting all the critical values $J_\perp$ for each given value of the Coulomb interaction $U$ of the conduction electrons. The system undergoes a phase transition from a Kondo singlet phase to a spiral magnetic order phase upon decreasing of $J_\perp$ and $U$. Representing the phase diagram in the $(J_\perp, ~J_\perp U)$ plane allows us to explain why Coulomb interaction $U$ disfavors the spiral magnetic order. Indeed, the theory of Kondo insulators \cite{fulde1993model} shows that the critical line between magnetism and Kondo singlet phases follows the equation:
\begin{eqnarray}\label{kondo_theory}
J^2_\perp + \alpha J_\perp U - \beta = 0,
\end{eqnarray}
which is a parabola like our phase diagram in Fig.~\ref{fig:SMOP-PD} (low panel). In Eq.\eqref{kondo_theory}, $\alpha$ is a constant and $\beta$ is generally a function of $J_\perp$ and $U$ and can be expanded as $1/(J_\perp + \alpha U)^2$. Fitting our numerical data, we found that $\alpha \approx 0.52t$ and $\beta \approx 4.12t$. These values are compatibles to the parameters $\alpha \approx 0.58t$ and $\beta \approx 4.26t$ found from square lattice in Ref.~\cite{faye2018phase}.
\begin{figure}
	\centerline{\includegraphics[width=9.0cm]{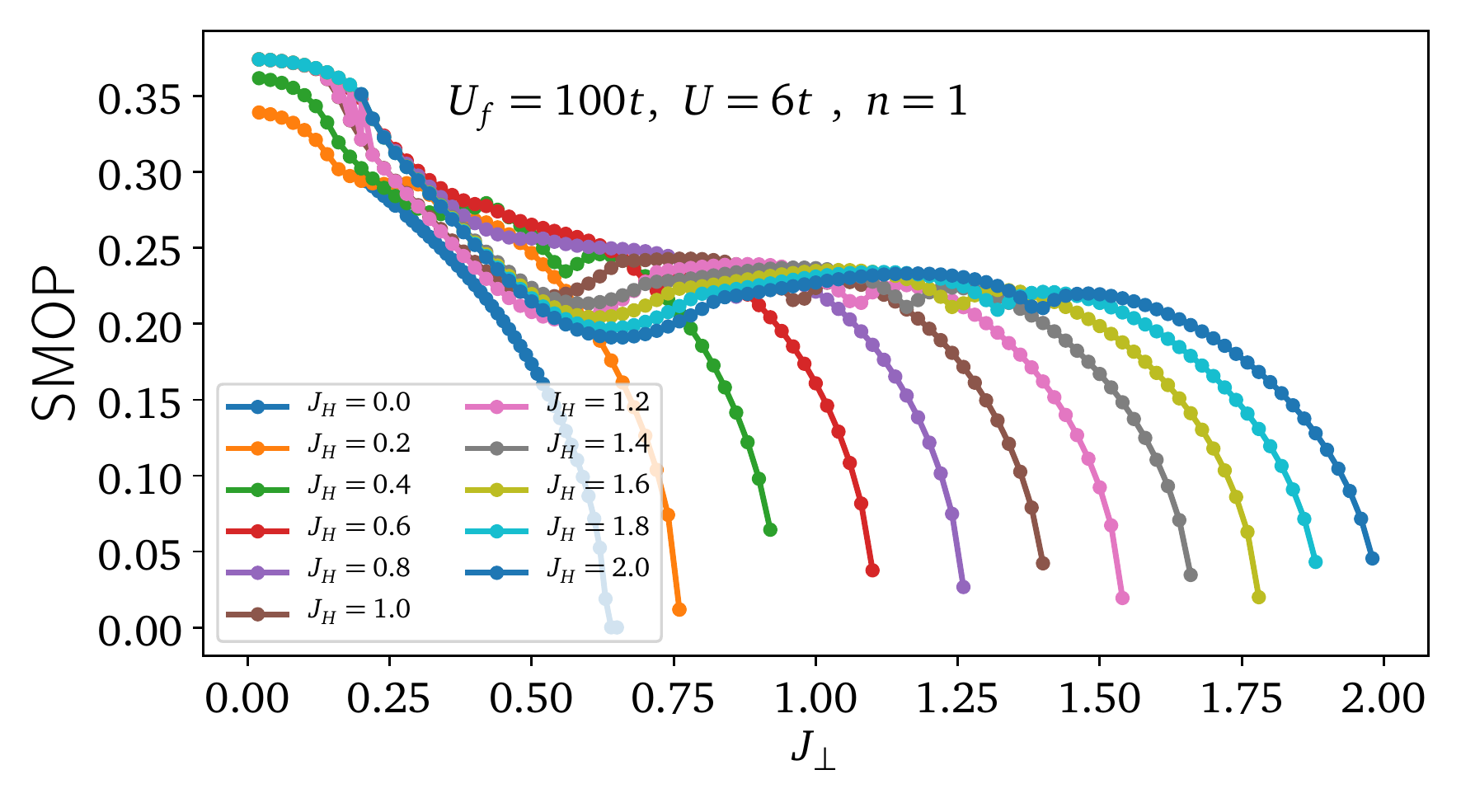}}
	\centerline{\includegraphics[width=9.0cm]{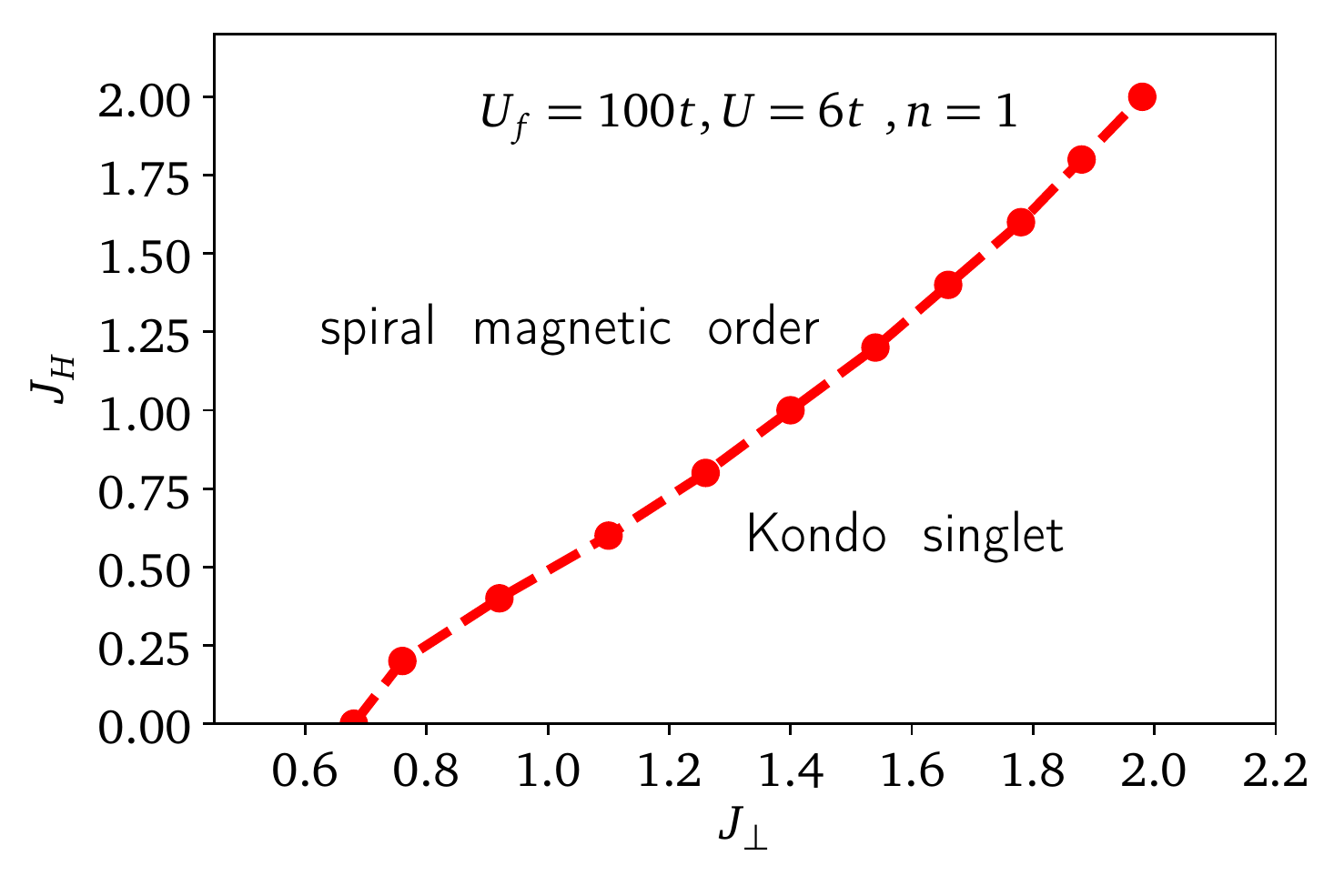}}
	\caption{Top panel: Spiral magnetic order parameter (SMOP) at half-filling and $U=6t$ for different values of localize spins Heisenberg exchange $J_H$. Low panel: Phase diagram in the plane $(J_H, J_{\perp})$ at $U=6t$.  We set $U_{f} = 100t$.} \label{fig:SMOP-PD-JH}
\end{figure}
In Fig.~\ref{fig:SMOP-PD-JH} (top panel), we show the spiral
magnetic order parameter at finite $J_H$ for $U = 6t$. We remark that the spiral magnetic order increases with
increasing $J_H$. In addition of that, the Heisenberg coupling $J_H$ also introduces some small quantum
oscillations of the spiral magnetic order parameter. These
quantum oscillations increase with increasing $J_H$. The theory of quantum oscillations of magnetization from
Ref.~\cite{PhysRevB.96.075115} predicted that the magnetization shows de Haas-van Alphen oscillations from
intermediate to weak Kondo coupling $J_\perp$. For the
Kondo-Hubbard triangular lattice model, these oscillations
are introduced by intermediate values of $J_H$.

In Fig.~\ref{fig:SMOP-PD-JH} (low panel), we present the quantum phase diagram at half-filling for $U=6t$ in the plane $(J_{\perp},J_H)$. Notice that at $J_H = 0$ and $U = 6t$, the critical spiral magnetic order parameter was found at $J_\perp = 0.68t$. The effect of the  exchange interaction $J_H$ is to push the spiral magnetic order into the Kondo singlet phase, given more room to the spiral magnetic phase. The transition from this spiral magnetic order to Kondo singlet phases is a second order phase transition and can be approximated with a straight line. Thus, even a large values of $J_{\perp}$ and $J_H$, the separation between the Kondo singlet and the spiral magnetic order phases is expected to exist. However, the exchange interaction $J_H$ must be limited to finite values (in general, small values compared to $J_\perp$) after which the system becomes a Kondo singlet phase at large $J_\perp$, as known from the theory \cite{ortmann2015topological, PhysRevB.51.15630, PhysRevLett.83.796, PhysRevB.64.092406, PhysRevB.56.11820}.

\subsection{Pure chiral superconductivity and competition with spiral magnetism}\label{away_hf}
In this section, we investigate $(i)$ how superconductivity is affected by the impurities via the coupling $J_\perp$, and $(ii)$ the competition between spiral magnetism and chiral superconductivity if the system is doped, i.e., away from half-filling. We fix the Coulomb interaction to $U=4t$ and the Heisenberg interaction to $J_H = 0$.
\subsubsection{Pure $d+id$ chiral superconductivity}
\begin{figure}
	\centerline{\includegraphics[width=9.0cm]{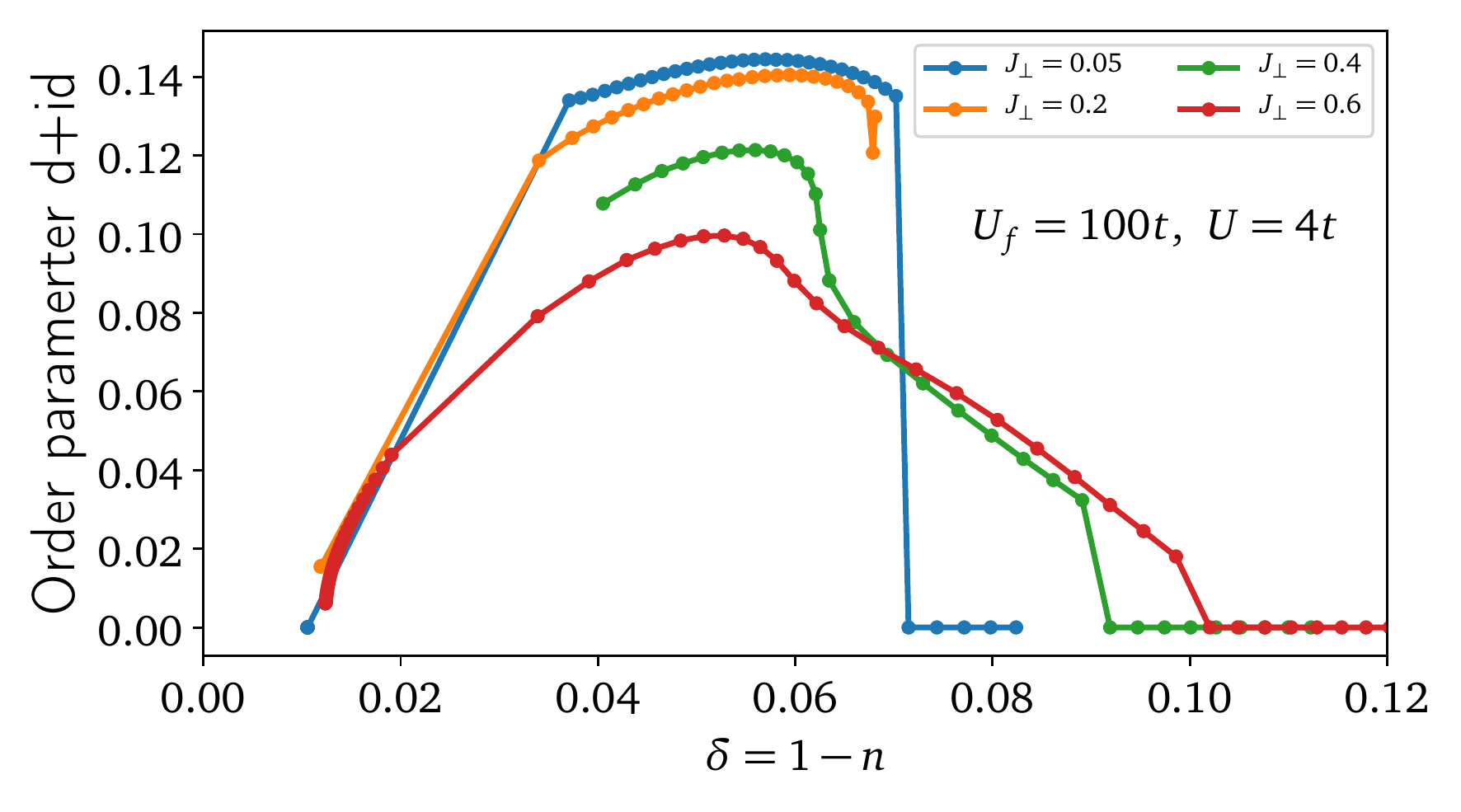}}
	\caption{Pure chiral singlet superconducting order parameter $d+id$ in function of hole doping $\delta = 1-n$ ($n$ is the electronic density of itinerant electrons) for different values of the coupling  $J_\perp$. We set the one site Coulomb interactions to $U=4t$ and $U_f = 100t$.} \label{fig:Di}
\end{figure}

In order to  prove broken symmetries in VCA, we need to add to the cluster the  Weiss fields corresponding to those broken symmetries. This is a requirement since VCA is a real-space method with an emphasis on short-range correlations because of the small size of the clusters. For superconductivity in particular, we introduce the following nearest-neighbor, singlet pairing operator: 
\begin{equation}
\label{eq:pairing}
\begin{aligned}
\hat{S}_{\rv,i} = c_{\rv,\uparrow}c_{\rv+\ev_i,\downarrow} - c_{\rv,\downarrow}c_{\rv+\ev_i,\uparrow},
\end{aligned}
\end{equation}
where $i=1, 2, 3$ and the nearest-neighbor vectors $\ev_{i}$ are the triangular lattice vectors. 
From this elementary operator, one can define a lattice-wide paring operator as follows:
\begin{equation}\label{eq:OP}
\hat{S}(\Qv) = \sum_ {\rv j}\left(\hat{S}_{\rv,j}e^{i(\Qv\cdot\rv +\phi_j)} + \mathrm{H.c.} \right).
\end{equation}
In Eq.~\eqref{eq:OP}, specifying the three phases $\phi_j$ and  the wave vector  $\Qv$ define the precise superconducting symmetry of the order parameter. In particular, the chiral singlet $d+id$ superconducting order is defined  for $(\phi_1,\phi_2,\phi_3) = (0, 2\pi/3, 4\pi/3)$ and  $\Qv=0$. In this configuration, all links have the same singlet pairing amplitude, but their phases vary. As we discuss in Sec. \ref{Sec:Model}, within Hubbard model, it was shown that for the two-dimensional triangular lattice, $d+id$ is favored among the possible superconducting pairings. Based on this, we will ignore here the other possible (singlet and triplet) superconducting pairings for the Kondo-Hubbard model. We only concentrate on the $d+id$ superconducting breaking symmetry. 
To investigate how impurities affect superconductivity via the coupling $J_\perp$, we also ignore the possibility of spiral magnetism for a moment.  We show in Fig.~\ref{fig:Di}, the pure chiral singlet superconducting order parameter $d+id$ in function of hole doping $\delta = 1-n$ for different values of $J_\perp$. 
We observe that the superconducting order parameter has a dome, like in high $T_c$ superconductors \cite{PhysRevLett.94.156404, PhysRevB.74.235117}, centered around $5\%$ doping. One remarks that the superconducting order parameter amplitude decreases with increasing coupling $J_\perp$. However, the range of doping where superconductivity exists increasing also with increasing $J_\perp$. Just, while the Kondo coupling reduces the amplitude of the superconducting order parameter, it helps $d+id$ to exist at moderate doping. We Notice a jump of the superconducting order parameter indicating that the transition from superconductivity to Kondo singlet phase is a first order phase transition. 
\subsubsection{Coexistence between spiral magnetism and chiral superconductivity}
\begin{figure}
	\centerline{\includegraphics[width=9.0cm]{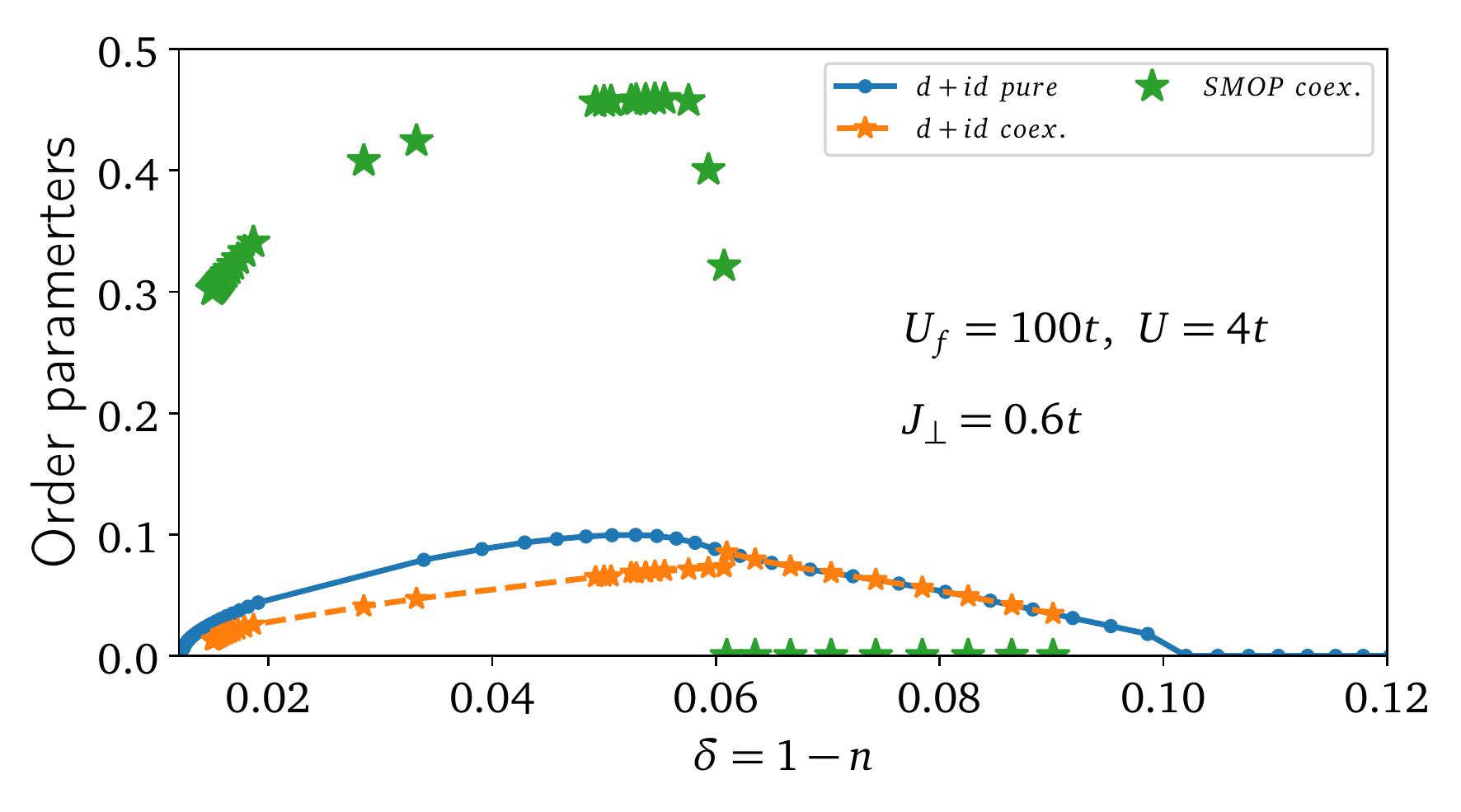}}
	\caption{The Spiral magnetic order parameter (SMOP), in green,  and the $d+id$ superconducting order parameter, in orange,  in a coexistence solution as a function of doping  $\delta = 1-n$ for $U = 4t$ and $J_{\perp} = 0.6t$. We also show, in blue, the individual pure $d+id$ superconducting order parameter.  We set $U_{f} = 100t$.} \label{fig:M-Di}
\end{figure}
To study the competition between spiral magnetism and chiral superconductivity, we add both the Weiss fields defined in Eqs. \eqref{op} and \eqref{eq:OP} in the cluster Hamiltonian, and let them vary simultaneously. Thus, this will correspond to a microscopic coexistence of spiral magnetism and chiral superconductivity if it is possible, otherwise the Weiss field of one of them will be zero.
We present in Fig.~\ref{fig:M-Di}, the spiral magnetic and $d+id$ superconducting order parameters in a coexistence solution as a function of doping for $U = 4t$ and $J_{\perp} = 0.6t$. We also show, in blue, the individual pure $d+id$ superconducting order parameter solution.
We observe a partial suppression in the amplitude of the superconducting order parameter due to the coexistence with spiral magnetism, indicating that these two phases are in competition. This suppression of superconductivity by the magnetism was also seen with singlet $d$-wave superconductivity on the square-lattice Hubbard model \cite{PhysRevLett.94.156404, PhysRevB.74.235117}. One the other hand, on honeycomb-lattice hubbard model, the triplet $p+ip$-wave was seen to be  enhanced by its coexistence with magnetism \cite{PhysRevB.93.155149}, which seems to be a cooperation between the two phases instead of competition  as we observe here.

\section{Conclusion}\label{conclusion}
Using the variational cluster approximation, we investigate the interplay between the quantum  magnetic order and Kondo singlet phase on one hand, and  the competition between the spiral magnetism and superconductivity on the other hand, within the Kondo-Hubbard triangular lattice. The latter includes the local Coulomb interaction $U$, the Kondo coupling $J_\perp$, and the Heisenberg exchange interaction $J_H$ of the impurities. At half-filling, we obtain the quantum ground state phase diagrams in the planes ($J_\perp, UJ_\perp$) and ($J_\perp, J_H$). In the ($J_\perp, UJ_\perp$) plane, the quantum phase diagram exhibits a spiral magnetic order phase at lower and intermediate $J_\perp$ and $U$, and a Kondo singlet phase at large $J_\perp$.  The transition from the spiral magnetic order phase to the Kondo singlet phase is found to be  a second order phase transition. In the ($J_\perp, J_H$) plane, we find that the effect of the Heisenberg $J_H$ is to push the magnetic order to high couplings $J_\perp$. It also introduces some small magnetic oscillations of the spiral magnetic order parameter. 

Away from half-filling, i.e., when the system is doped, we study the effects of Kondo coupling on pure superconductivity and the competition of the latter phase with spiral magnetism. In absence of the spiral  magnetism,  we find that, the range of doping where the $d+id$-wave  superconductivity is found increases with moderate couplings $J_\perp$. However, the amplitude of the superconducting order parameter decreases with increasing of the coupling. In presence of the spiral magnetism, we observe a competition between this spiral magnetism and the superconductivity, which results in partial suppression of the superconducting order parameter amplitude . This competition between magnetism and superconductivity was also observed in the Hubbard square lattice model, which describes high-$T_c$ cuprates, where the  antiferromagnetism was found to suppress the amplitude of the $d$-wave superconducting order parameter.
\begin{acknowledgments}
We gratefully acknowledge conversations with D. S\'en\'echal and M. Kiselev. Computing resources were provided by Compute Canada and Calcul Qu\'ebec. 
\end{acknowledgments}

\bibliography{papier}
\end{document}